\documentclass[a4paper,twocolumn,11pt,accepted=2025-06-25]{quantumarticle}
\pdfoutput=1
\usepackage[numbers,sort&compress]{natbib}
\usepackage[utf8]{inputenc}
\usepackage[english]{babel}
\usepackage[T1]{fontenc}
\usepackage{enumerate}
\usepackage{graphicx}
\usepackage{caption}
\usepackage{subcaption}
\usepackage{dcolumn}
\usepackage{bm}
\usepackage[unicode, colorlinks=true,linkcolor=blue,citecolor=blue,urlcolor=blue]{hyperref}
\usepackage{amsthm}
\usepackage{amsmath}
\usepackage{algorithm2e}
\RestyleAlgo{ruled} 

\usepackage{dsfont}
\usepackage{xcolor}
\usepackage{ragged2e}
\usepackage[noend]{algpseudocode}
\usepackage{booktabs}

\usepackage{braket}
\newcommand{\tr}[1]{\mathrm{Tr}\bigl(#1\bigr)}
\usepackage{csquotes}
\usepackage{bbold}

\begin{document}

\title{Sampling Groups of Pauli Operators to Enhance Direct Fidelity Estimation}

\author{Júlia Barberà-Rodríguez}

\affiliation{ICFO - Institut de Ciencies Fotoniques, The Barcelona Institute of Science and Technology, 08860 Castelldefels, Barcelona, Spain}
\orcid{0009-0004-8445-3345}

\author{Mariana Navarro}

\affiliation{ICFO - Institut de Ciencies Fotoniques, The Barcelona Institute of Science and Technology, 08860 Castelldefels, Barcelona, Spain}
\affiliation{LuxQuanta Technologies S.L., Mediterranean Technology Park. Carrer d’Esteve Terradas, 1, Office 206, 08860 Castelldefels, Barcelona, Spain}
\orcid{0000-0001-9381-369X}

\author{Leonardo Zambrano}
\email{leonardo.zambrano@icfo.eu}
\affiliation{ICFO - Institut de Ciencies Fotoniques, The Barcelona Institute of Science and Technology, 08860 Castelldefels, Barcelona, Spain}
\orcid{0000-0001-7070-1433}

\begin{abstract}
Direct fidelity estimation is a protocol that estimates the fidelity between an experimental quantum state and a target pure state. By measuring the expectation values of Pauli operators selected through importance sampling, the method is exponentially faster than full quantum state tomography. We propose an enhanced direct fidelity estimation protocol that uses fewer copies of the experimental state by grouping Pauli operators before the sampling process. We derive analytical bounds on the measurement cost and estimator variance, showing improvements over the standard method. Numerical simulations validate our approach, demonstrating that for 8-qubit Haar-random states, our method achieves a one-third reduction in the required number of copies and reduces variance by an order of magnitude using only local measurements. These results underscore the potential of our protocol to enhance the efficiency of fidelity estimation in current quantum devices.
\end{abstract}

\maketitle

\section{Introduction \label{sec:intro}}

The rapid advancements in quantum technologies have underscored the need for efficient methods to characterize quantum devices.
Accurate characterization plays a crucial role in verifying hardware performance, diagnosing errors, and ensuring the reliability of quantum protocols. Among various characterization tasks, fidelity estimation is of particular importance, as it quantifies how closely a prepared quantum state matches a desired target state. Traditionally, this verification has relied on quantum state tomography (QST)~\cite{james2001measurement, thew2002qudit}, which reconstructs the full quantum state of a physical system. However, QST is computationally expensive, requiring at least $\mathcal{O}(4^n)$ copies of the experimental state, where $n$ is the number of qubits of the system. It also demands substantial classical resources for computation and storage, making it impractical for large quantum systems.

For fidelity estimation specifically, more targeted and efficient techniques can help address these challenges. One such method is direct fidelity estimation (DFE), which involves measuring the expectation values of Pauli operators selected via importance sampling~\cite{flammia2011direct, da2011practical, fawzi2024optimalfidelityestimationbinary}. DFE offers a practical and scalable approach for fidelity estimation, relying exclusively on local measurements and being exponentially faster than QST. However, in the worst case, the number of measurements required by DFE  scales as $\mathcal{O}(2^n)$. Another promising approach are classical shadows (CS), which employ random measurements to estimate the expectation values of observables~\cite{huang2020predicting}. Although CS are effective for a variety of tasks, the scaling for fidelity estimation of arbitrary states is at least $\mathcal{O}(3^n)$ when constrained to local measurements. To mitigate this, a derandomized version of CS has been proposed, which reduces the measurement overhead for arbitrary observables~\cite{huang2021efficient}.  Nevertheless, for the specific task of fidelity estimation with local measurements, DFE remains one of the most effective and practical methods available.

While DFE provides a more efficient framework for fidelity estimation, its practical implementation involves measuring a large number of Pauli operators and copies of the experimental state. This poses significant challenges for systems with many qubits, particularly in the context of noisy intermediate-scale quantum (NISQ) devices where measurement resources are limited~\cite{RevModPhys.94.015004}. To address these challenges, improved measurement strategies are necessary to optimize the extraction of information from each experimental setting. A widely studied solution involves grouping Pauli operators that commute, enabling them to be measured simultaneously within a single experimental setting~\cite{kandala2017hardware}. Such grouping techniques, commonly used in estimating Hamiltonian expectation values, significantly reduce the number of measurements required without sacrificing accuracy. Over the past decade, various methods have been proposed to optimize Pauli operator groupings, leading to substantial improvements in the scalability of operator estimation~\cite{izmaylov2019revising, verteletskyi2020measurement, jena2022optimization, yen2020measuring, izmaylov2019unitary, zhao2020measurement, hamamura2020efficient, Crawford_2021, yen2023deterministic}.
In addition to their use in Hamiltonian expectation value estimation, grouping techniques have also been applied to arbitrary observable estimation within the framework of classical shadows~\cite{hadfield2022measurements, gresch2023guaranteed, wu2023overlapped}. However, these methods are primarily designed for general tasks and have not been particularly optimized for fidelity estimation.

In this work, we propose a method to reduce the number of measurements required in DFE based on Pauli grouping. Instead of evaluating individual Pauli strings, we group them into commuting families and perform weighted sampling over these groups. This approach allows us to derive tighter bounds on the number of copies of the state required for the protocol. Additionally, we prove that the variance of our estimator is upper bounded by the variance of the original DFE estimator under certain conditions. Our theoretical analysis demonstrates that this optimized protocol outperforms the standard DFE in terms of both measurement cost and statistical efficiency. We validate our theoretical results through numerical simulations, exploring various grouping techniques.  The simulations reveal that even when restricted to local measurements, our method significantly reduces the number of copies required by one-third while achieving a reduction of one order of magnitude in the variance compared to the original DFE protocol. 

This paper is structured as follows. Section~\ref{sec:preliminaries} provides an overview of the DFE protocol and existing Pauli grouping techniques. In Section~\ref{sec:method}, we detail our optimized approach and present variance and measurement cost bounds. Section~\ref{sec:numerical} discusses the results of numerical simulations, with conclusions drawn in Section~\ref{sec:disc}.

\section{Preliminaries \label{sec:preliminaries}}

\subsection{Direct fidelity estimation\label{sec:dfe}}
    Assume we have a $n$-qubit state $\sigma$ prepared in a physical system and a target state $\rho$. The protocol of direct fidelity estimation (DFE) allows us to estimate the fidelity $F(\rho, \sigma) = \tr{\rho \sigma}$, given that $\rho$ is a pure state \cite{flammia2011direct}. Both $\rho$ and $\sigma$ states can be expressed in terms of their Pauli expectation values as 
    \begin{equation}
        \sigma  = \sum_{k} \frac{a_k }{\sqrt{d}} P_k, \qquad  \rho  =  \sum_{k} \frac{b_k}{\sqrt{d}} P_k. \label{eq:state in pauli basis}
    \end{equation}
    Here, each $P_k$ is a tensor product of $n$-qubit Pauli operators or the identity $\{\sigma_x, \sigma_y, \sigma_z, \mathds{1}\}$, with associated coefficients $a_k = \tr{\sigma P_k/\sqrt{d}}$ and $b_k$ defined analogously for $\rho$, where $d = 2^n$ is the dimension of the Hilbert space. Then, the fidelity between these two states reads 
    \begin{align}
        F(\rho,\sigma) = \sum_k a_k b_k = \sum_k b_k^2\frac{a_k }{b_k }, \label{eq: fidelidad original}
    \end{align}
    where the sum runs over all $k$ for which $b_k \neq 0$.
    Since the target state $\rho$ is pure, it is satisfied that $\sum_k b_k^2 = 1$. Consequently, the fidelity can be interpreted as the expected value of a random variable $X$, with values   
    \begin{align}
        X_k = \frac{a_k}{b_k} = \frac{\tr{\sigma P_k/\sqrt{d}}}{ \tr{\rho P_k/\sqrt{d}}} \label{eq: original estimator DFE}
    \end{align}
    and probability distribution $\Pr(X =X_k) = b_k^2$. 
    
    By applying Monte Carlo importance sampling to the probability distribution $b_k^2$, we can estimate the observable $X_k$ with additive error $\epsilon$ and failure probability $\delta$. The steps to achieve this are outlined below, which form the core of the DFE protocol~\cite{flammia2011direct, Kliesch2021ReviewCertification}:
    \begin{enumerate}
        \item Generate $\ell$ independent and identically distributed random samples $k_1,..., k_\ell$ from the importance sampling distribution $b^2_{k_i}$, where $i \in \{1,..,\ell\}$ and $\ell := \lceil (\epsilon^2 \delta)^{-1}\rceil$.
        \item Measure each observable $P_{k_i}$ a number of $m_{k_i}$ times, where $m_{k_i}$ is defined as
        \begin{equation}\label{eq:copies-original}
            m_{k_i} := \left \lceil \frac{2}{ b^2_{k_i} d \ell \epsilon^2 } \ln{\frac{2}{\delta}} \right\rceil .
        \end{equation}
        \item For each $i$, calculate the empirical estimate of the expectation value $\tr{\sigma P_{k_i}}$ based on the measurements outcomes from step 2. Using these estimates, calculate the estimator $\hat{X}_{k_i}$ for $X_{k_i} := a_{k_i}/b_{k_i}$.
        \item Calculate $\hat{Y} := \frac{1}{\ell} \sum^\ell_{i=1} \hat{X}_{k_i}.$
        \item Return $\hat{Y}$ as a fidelity estimator.
    \end{enumerate}
    
    The fidelity estimate $\hat{Y}$ serves as an unbiased estimator of $F(\rho,\sigma)$ with an accuracy of $2\epsilon$ and a confidence level of $1-2\delta$. The protocol guarantees that the expected number of state preparations $m_{k_i}$ is bounded and scales at most linearly with the Hilbert space dimension.

\subsection{Grouping Pauli operators\label{sec:grouping}}

Any $n$-qubit quantum state $\rho$ can be expressed as a linear combination of Pauli strings as shown in Eq.~\eqref{eq:state in pauli basis}. These strings can be grouped into sets of mutually commuting elements, such that 
\begin{equation}
\rho =  \sum_{k=1}^g \sum_{l=1}^{g_k} a_{kl} P_{kl},
\end{equation}
where $g$ denotes the total number of groups, $P_{kl}$ represents the Pauli strings within the $k$-th group, $a_{kl}$ are their corresponding weights, and $g_k$ is the number of Pauli strings in the $k$-th group. This grouping enables all operators within each set to be measured on a common basis, thereby reducing the number of experimental settings required and simplifying the measurement process~\cite{yen2023deterministic}.

\begin{figure}[ht]
    \includegraphics[width=\columnwidth]{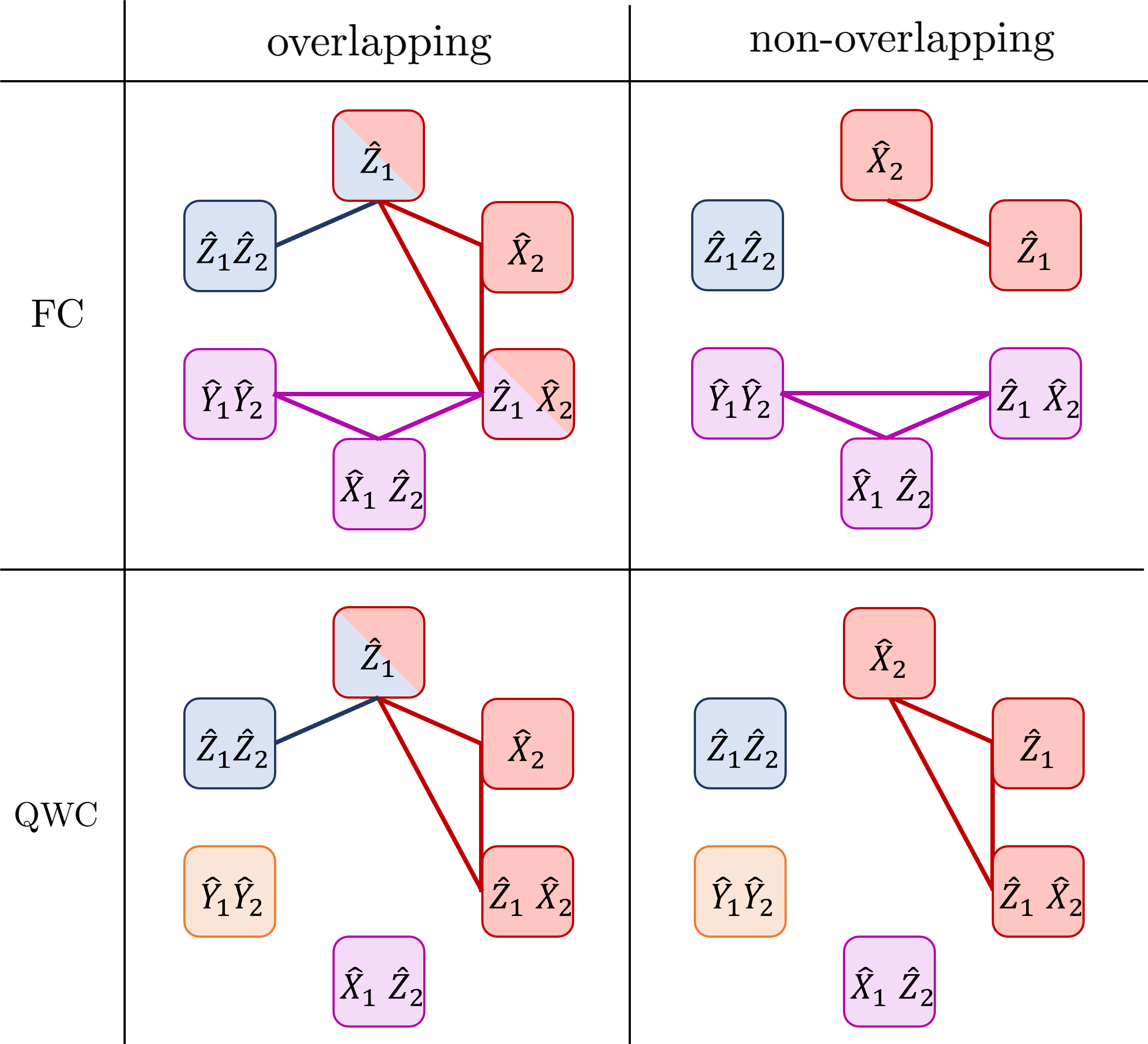}
    \caption{\justifying Diagram illustrating the different frameworks for grouping Pauli operators into commuting families. The x-axis distinguishes between overlapping groupings, where Pauli operators can belong to multiple groups, and non-overlapping groupings, where each Pauli operator is assigned to a single group. The y-axis differentiates between qubit-wise commutativity (QWC) and full commutativity (FC). Pauli strings in the same group share a color and are linked by solid lines.  }
    \label{fig:grouping-diagram-1}
\end{figure}

There are two key aspects to consider in the grouping process. The first involves selecting an appropriate commutativity framework, with two primary approaches commonly employed. The full commutativity (FC)~\cite{yen2020measuring} technique groups Pauli strings when the associated operators commute, that is, $\left[P_k, P_l\right] = \left[\otimes_{j=1}^n \sigma_{kj}, \otimes_{j=1}^n \sigma_{lj}\right] = 0$. On the other hand, qubit-wise commutativity (QWC)~\cite{verteletskyi2020measurement} groups Pauli strings only if they commute at the level of single-qubit operators. Specifically, $P_k$ and $P_l$ can belong to the same group if $\left[\sigma_{kj}, \sigma_{lj}\right] = 0$ for all $j = 1, \dots, n$. Although, in general, FC reduces the number of measurements needed and the variance of estimators compared to QWC, the implementation of FC requires two-qubit gates to perform the necessary basis changes for simultaneous measurements that increase the experimental overhead~\cite{yen2021cartan, Crawford_2021, yen2023deterministic}.

The second consideration is whether to allow Pauli operators to appear in multiple groups. Initial methods, such as the sorted insertion (SI) algorithm~\cite{Crawford_2021}, followed a non-overlapping framework, where each Pauli operator is assigned to a single group. Using these non-overlapping constraints and operating under QWC, SI minimizes variance by sorting Pauli operators by their weights in descending order before grouping them. While SI is simple, overlapping frameworks have shown improved performance by allowing Pauli operators to appear in multiple groups~\cite{wu2023overlapped, shlosberg2023adaptive}. Methods like iterative coefficient splitting (ICS) and iterative measurement allocation (IMA)~\cite{yen2023deterministic} exploit overlaps to further reduce the number of measurements by redistributing coefficients among groups based on the covariance between commuting Pauli operators.
Although ICS and IMA offer better performance in terms of reducing measurements and variance, ICS and IMA rely on SI to first generate non-overlapping groups, followed by additional steps for overlapping. This introduces extra computational costs, reducing their overall efficiency compared to SI. An illustrated comparison of the different grouping frameworks can be seen in Fig. \ref{fig:grouping-diagram-1}.

\section{Method \label{sec:method}} 
    
    Considering an $n$-qubit system, the fidelity in Eq.~\eqref{eq: fidelidad original} requires evaluating the expectation values of up to $4^n$ Pauli strings. This number can be significantly reduced by grouping the Pauli strings into $g$ groups of mutually commuting operators, which can be measured simultaneously. For $k = 1, 2, \dots, g$, the $k$-th group is defined as $\{P_{k1}, P_{k2}, \dots P_{k g_k}\}$,
    where $g_k$ is the number of elements in the group. For states $\sigma$ and $\rho$ we define the vectors $\textbf{a}_k = (a_{k1}, a_{k2}, \dots, a_{k g_k})$ and $\mathbf{b}_k = (b_{k1}, b_{k2}, \dots, b_{k g_k})$, where $a_{kl}= \tr{\sigma P_{kl}/\sqrt{d}}$ and $b_{kl}= \tr{\rho P_{kl}/\sqrt{d}}$. Using this grouping, the fidelity between the target pure state $\rho$ and the experimental state $\sigma$ can be expressed as
    \begin{align}\label{eq:fidelity_grouped}
        F(\rho,\sigma) =  \sum_{k=1}^g \mathbf{a}_{k} \cdot \mathbf{b}_{k} =  \sum_{k=1}^g \lVert \mathbf{b}_k \rVert^2  \frac{\mathbf{a}_{k} \cdot \mathbf{b}_{k}}{\lVert \mathbf{b}_k \rVert^2} ,
    \end{align}
     where $\mathbf{a}_{k}  \cdot \mathbf{b}_{k} = \sum_{l=1}^{g_k} a_{kl} b_{kl}$ and
     \begin{align}
         \lVert \mathbf{b}_k \rVert = \left(\sum^{g_k}_{l=1} b_{kl}^2\right)^{1/2} \label{eq: new prob dist}
     \end{align}
    is the Euclidian norm of $\mathbf{b}_k$. Since $\rho$ is a pure state, $\sum_{k=1}^g \lVert \mathbf{b}_k \rVert^2 =1$.

    Building on the original DFE protocol outlined in Sec.~\ref{sec:dfe}, we introduce a modified probability distribution $\lVert \mathbf{b}_k \rVert^2$, determined by the coefficients of the grouped Pauli strings associated with the target pure state $\rho$. Consequently, the values $X_k$ in Eq. \eqref{eq: original estimator DFE} of the random variable $X$ are redefined as
    \begin{align}
        X_k =\frac{\mathbf{a}_{k}  \cdot \mathbf{b}_{k}}{\lVert \mathbf{b}_k \rVert^2} = \sum_{l=1}^{g_k} \frac{a_{kl} b_{kl}}{\sum_{j=1}^{g_k} b_{kj}^2}.
    \end{align}
   The exact value of $X_k$ can be determined under the assumption that $\mathbf{a}_k$ is estimated perfectly, using as many copies of $\sigma$ as necessary. 

    \begin{figure*}[ht]
    \centering
    \includegraphics[width=1.8\columnwidth]{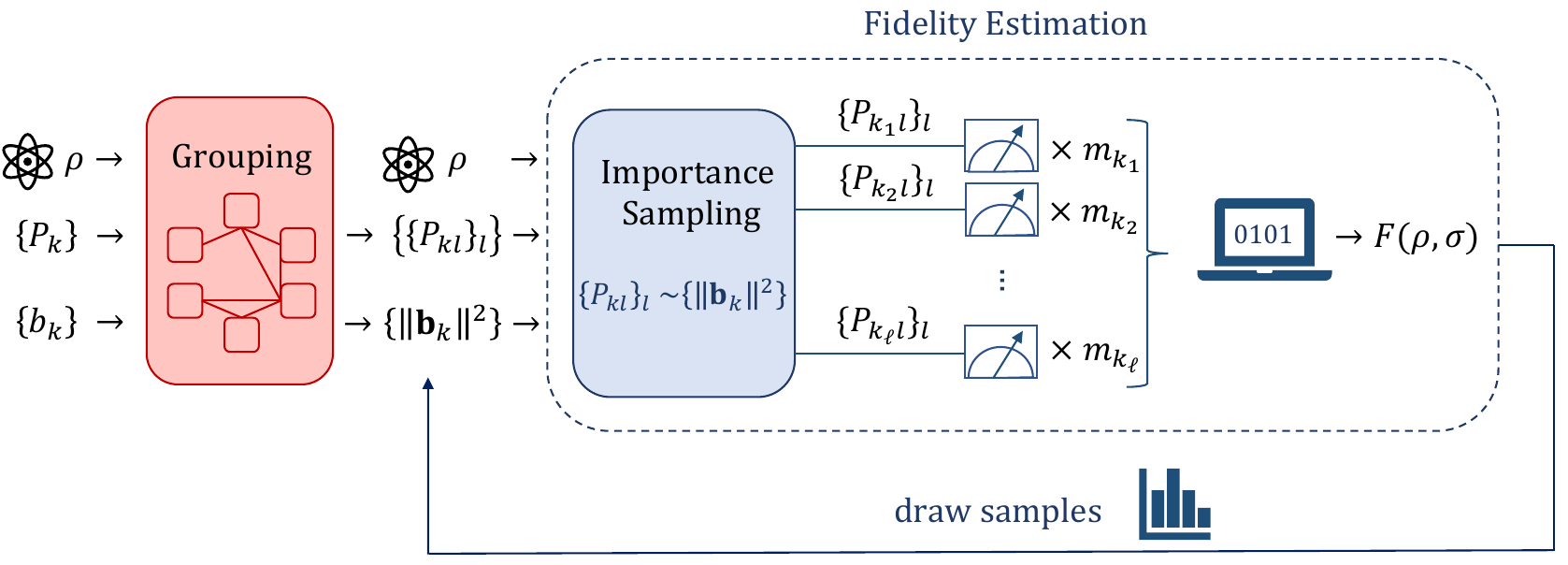}
    \caption{\justifying Improved protocol for estimating the fidelity between an unknown experimental state $\sigma$ and a pure state $\rho$. The algorithm inputs include the $n$-qubit target state $\rho$, a list of $n$-qubit Pauli strings $\{ P_{k} \}$, and their corresponding weights $\{ b_{k} \}$. First, the Pauli strings are grouped into commuting families $\{ \{P_{{k}l}\}_l \}$, and the new weights $\{ \lVert\mathbf{b}_{k}\rVert^2 \}$ are calculated using the chosen technique. The DFE protocol is then performed, using the groups as elements for the sampling process with the probability distribution $\{ \lVert \mathbf{b}_{k} \rVert^2 \}$. Each sampled group of Pauli strings is measured $m_{k_i}$ times, requiring a total of $\sum_{k_i} m_{k_i}$ copies of the target state $\rho$.  }
    \label{fig:grouping-diagram}
    \end{figure*}

    To apply Monte Carlo importance sampling with the probability distribution $P(X =X_k) = \lVert \mathbf{b}_k \rVert^2$, we would like to satisfy two conditions. First, we demand $X$ to be an unbiased estimator of $\mathrm{Tr}(\rho \sigma)$ with respect to the probability distribution $\lVert \mathbf{b}_k \rVert^2$, that is, $\mathbb{E}_{X \sim \lVert \mathbf{b}_k \rVert^2} [X] = \mathrm{Tr}(\rho \sigma)$. This follows directly from Eq.~\eqref{eq:fidelity_grouped}. The second condition is that the variance of the estimator is bounded. This can be seen from  
    \begin{align}
        \text{Var}(X) &= \sum^g_{k=1} \lVert \mathbf{b}_k \rVert^2 \left(  \frac{\textbf{a}_{k} \cdot \mathbf{b}_{k}}{\lVert \mathbf{b}_k \rVert^2}\right)^2 - \tr{\rho \sigma }^2 \nonumber \\
        & \leq \sum^g_{k=1} \frac{1}{\lVert \mathbf{b}_k \rVert^2} \lVert \mathbf{a}_k \rVert^2 \lVert \mathbf{b}_k \rVert^2 \nonumber \\
        & \leq 1,
    \end{align}
    where in the second line we have used the Cauchy–Schwarz inequality.

    We aim to estimate $F(\rho, \sigma)$ with additive error $2 \epsilon$ and confidence $1 - 2 \delta$. To achieve this, we generate $\ell = \lceil (\epsilon^2 \delta)^{-1} \rceil$ independent and identically distributed samples ${k_1, \dots, k_\ell}$ according to the probability distribution $\lVert \mathbf{b}_k \rVert^2$. Each $k_i$ corresponds to an independent realization $X_{k_i}$ of the random variable $X$. Assuming that all the $X_{k_i}$ can be known perfectly, the empirical mean is
    \begin{align}
    Y = \frac{1}{\ell} \sum_{i=1}^\ell X_{k_i}.
    \end{align}
    By applying Chebyshev's inequality, we have that $Y$ satisfies
    \begin{align} \label{eq:theoretical bound Y-F()}
    \mathrm{Pr}(|Y - \mathrm{Tr}(\rho \sigma)| \geq \epsilon) \leq \frac{1}{\ell \epsilon^2}.
    \end{align}
    Thus, the estimator $Y$ is $\epsilon$-close to the fidelity $F(\rho, \sigma)$ with probability at least $1 - \delta$, for any arbitrarily small $\delta$.

    In a more realistic setting, the ideal estimator $Y$,  which assumes infinite precision and unlimited resources, is approximated by a finite-sample estimator $\tilde{Y}$  that uses a limited number of copies of the state $\sigma$. Then we need to establish a bound for the approximation error $| \tilde{Y} - Y |$ using Hoeffding's inequality.  
    For this, we perform an eigendecomposition of the Pauli operators $P_{kl}$ within group $k$. Since all the operators in each group commute, they share a common eigenbasis, denoted by $\{ |r_k\rangle \}_{r=1}^{d}$. Then,
    \begin{align}
        P_{kl} = \sum^{d}_{r = 1} c_{kl}^{(r)} |r_{k} \rangle \langle r_{k}|,
    \end{align}
    where $\{c_{kl}^{(r)}\} \in \{-1,1\}$ are the corresponding eigenvalues.  
    
    For any realization $k_1, \dots, k_\ell$ of the experiment, the value $k_i$ indicates that the group of Pauli strings $\{ P_{k_i l} \}_{l=1}^{g_{k_i}}$ must be measured in the $i$-th round of the experiment. We prepare $m_{k_i}$ (defined below) copies of the state $\sigma$, and measure them using the POVM $\{|r_{k_i} \rangle \langle r_{k_i}|\}_{r=1}^d$. For each measurement $j=1, \dots m_{k_i}$, we obtain an outcome $r_{k_{i} j}$ with probability $p(r_{k_i j}|k_i) = \langle r_{k_i}| \sigma |r_{k_i} \rangle$.  Then, we define the estimator $\tilde{X}_{k_{i}}$ of $X_{k_i}$ as 
    \begin{align}
        \tilde{X}_{k_{i}} = \frac{1}{m_{k_i} \lVert \mathbf{b}_{k_i} \rVert^2 \sqrt{d}} \sum_{j= 1}^{m_{k_i}} C_{{k_{i}} r_j}, \label{eq:precise estimator}
    \end{align}
     where $C_{{k_{i}} r_j } = \sum_l c_{{k_{i}} l}^{(r_j)} b_{{k_{i}}l}$ and $c_{k_i l}^{(r_j)}$ denotes the $r_j$-th eigenvalue of the Pauli operator $P_{k_i l}$ obtained during the $j$-th round of the experiment. The expectation value of the measurement outcomes $c_{k_i l}^{(r_j)}$ is
    \begin{align}
        \mathbb{E} \left[ c_{{k_{i}} l}^{(r_j)} \right] =& \sum^{d}_{r = 1} c_{k_il}^{(r)} \langle r_{k_i} | \rho | r_{k_i} \rangle \nonumber \\
        =&   \mathrm{Tr} \left(\sigma P_{k_il} \right) \nonumber\\
        =&  \sqrt{d} \;a_{k_i l} .
    \end{align}
    Then, we have $\mathbb{E}\left[C_{{k_{i}} r_j }\right] = \sqrt{d} \sum_l a_{k_i l} b_{k_i l} = \sqrt{d} \mathbf{a}_{k_i} \cdot \mathbf{b}_{k_i} $ and since expectation values are linear, $\tilde{X}_{k}$ is an unbiased estimator of $X_k$.

    We now define an unbiased estimator for the random variable $\tilde{Y}$ as follows
    \begin{align}\label{eq:our-estimator}
         {\ell} \tilde{Y} &= \sum_{i=1}^\ell \tilde{X}_{k_{i}} \nonumber \\
        & =  \sum_{i=1}^\ell \sum_{j=1}^{m_{k_i}}  \frac{ C_{k_{i} r_j}}{m_{k_i}  \lVert \mathbf{b}_{k_i} \rVert^2 \sqrt{d}}.
    \end{align}

    Using Hölder's inequality, we observe that $|C_{{k_{i}} r_j } |\leq \lVert \mathbf{b}_{k_i} \rVert_1  \max_l |c_{k_i l}^{(r_j)}|  = \lVert \mathbf{b}_{k_i} \rVert_1$, which provides a bound for the argument of the sum in Eq. \eqref{eq:our-estimator}:
    \begin{align}
        \left|\frac{ C_{k_{i} r_j}}{m_{k_i}  \lVert \mathbf{b}_{k_i} \rVert^2 \sqrt{d}} \right| \leq   \frac{ \lVert \mathbf{b}_{k_i} \rVert_1 }{m_{k_i} \lVert \mathbf{b}_{k_i} \rVert^2 \sqrt{d}} \label{eq:bounds empirical estimator}.
    \end{align}
    Applying Hoeffding's inequality for $ \epsilon \ell > 0 $ and using the bounds from Eq.~\eqref{eq:bounds empirical estimator}, we obtain the following inequality
    \begin{align}
        &\Pr(| \tilde{Y} - Y| \geq \epsilon) \nonumber\\
        &\leq 2 \exp \left(\frac{-2\epsilon^2 \ell^2}{\sum_{i=1}^\ell \sum^{m_{k_i}}_{j=1}   \left(\frac{2 \lVert \mathbf{b}_{k_i} \rVert_1}{m_{k_i}  \lVert \mathbf{b}_{k_i} \rVert^2 \sqrt{d}}  \right)^2} \right) \nonumber\\
        & =2 \exp \left(\frac{-\epsilon^2 \ell^2}{\sum_{i=1}^\ell  \frac{2  \lVert \mathbf{b}_{k_i} \rVert_1^2}{m_{k_i}  \lVert \mathbf{b}_{k_i} \rVert^4 d}}  \right).
    \end{align}
    We want this probability to be smaller than a given $\delta$, that is,
    \begin{align}
        \Pr(| \tilde{Y} -  Y| \geq  \epsilon) \leq \delta \label{eq: bounded prob by delta}.
    \end{align}
    For this to hold, we need 
    \begin{align}
        \ln{\frac{2}{\delta}} \leq \frac{\epsilon^2 \ell^2}{ \sum_{i=1}^\ell  \frac{ 2  \lVert \mathbf{b}_{k_i} \rVert_1^2}{m_{k_i}   \lVert \mathbf{b}_{k_i} \rVert^4 d}},
    \end{align}
    from which we obtain the number of copies of the state needed for each group,
    \begin{align}\label{eq:copies-our}
        m_{k_i} &= \left\lceil \frac{2 \lVert \mathbf{b}_{k_i} \rVert_1^2}{  \lVert \mathbf{b}_{k_i} \rVert^4 d  \ell  \epsilon^2 } \ln{\frac{2}{\delta}} \right\rceil.
    \end{align} 

    We have established a bound on the approximation error $|Y - \tilde{Y}|$. By combining this result with the bound in Eq. \eqref{eq:theoretical bound Y-F()} and using the union bound, we can conclude that
    \begin{align}
        \Pr[\tilde{Y}-F(\rho, \sigma) \leq 2 \epsilon] \geq 1-2\delta.
    \end{align}
    Thus, when employing our protocol (see Fig. \ref{fig:grouping-diagram}), the fidelity is guaranteed to be within the interval $[\tilde{Y}-2\epsilon, \tilde{Y}+2\epsilon]$ with confidence $1-2\delta$.

    Let us now calculate a bound for the average number of copies in our protocol. We sample $\ell = \lceil 1/(\epsilon^2 \delta) \rceil$ groups, with each group measured $m_{k_i}$ times. Then, the average number of copies per group is bounded by
    \begin{align}\label{eq:expected_mi}
        \mathbb{E}[m_{k_i}] & \leq 1 + \sum_{k=1}^{g} \frac{2 \lVert \mathbf{b}_{k_i} \rVert_1^2}{ \lVert \mathbf{b}_{k_i} \rVert^2d \ell \epsilon^2}\ln{\frac{2}{\delta}}.
    \end{align}
    
    The total number of copies of the state is $m = \sum_{i=1}^\ell m_{k_i}$. Therefore, for the expected total number of copies of our protocol, we have 
    \begin{align}
        \mathbb{E}\left[\sum^\ell_{i=1} m_{k_i}\right] 
        &\leq 1 + \frac{1}{\epsilon^2 \delta} + \sum_{k=1}^{g} \frac{2 \lVert \mathbf{b}_{k_i} \rVert_1^2}{ \lVert \mathbf{b}_{k_i} \rVert^2d \epsilon^2}\ln{\frac{2}{\delta}}.
    \end{align}
    Since $ \lVert \mathbf{b}_{k_i} \rVert_1 \leq \sqrt{g_{k_i}}  \lVert \mathbf{b}_{k_i} \rVert$ and $\sum_{k=1}^{g} g_{k_i} $ is at most $d^2$ for an arbitrary grouping technique, we find that the total number of copies for our protocol is bounded by that of the original DFE protocol:
    \begin{align}
        \mathbb{E}\left[\sum^\ell_{i=1} m_{k_i}\right] &\leq  1 + \frac{1}{\epsilon^2 \delta} + \frac{2}{d \epsilon^2} \ln{\frac{2}{\delta}}\sum_{k=1}^{g} g_{k_i} \nonumber \\
        & \leq 1 + \frac{1}{\epsilon^2 \delta} + \frac{2d}{\epsilon^2} \ln{\frac{2}{\delta}}.
    \end{align}

    Moreover, in App.~\ref{app:variance-bound}, we show that the variance of our estimator, $\mathrm{Var}(\tilde{Y})$, is upper bounded by the variance of the original DFE estimator, $\mathrm{Var}(\hat{Y})$, when the coefficients satisfy $\Vert \mathbf{b}_{k} \Vert_1^2 / \Vert \mathbf{b}_{k} \Vert^4 \gg 1$ (e.g. for Haar-random states when $d$ is large, since $|b_{kl}| \approx 1/d$). In this regime, $m_{k} \gg 1$, and the approximation $\lceil m_{k} \rceil \approx m_{k}$ holds.  Consequently, the fidelity estimates obtained with our protocol are more accurate than those from the original protocol. 
    

   Fidelity estimation can also be performed using shadow-based methods, such as classical shadows and their derandomized variant \cite{huang2020predicting, huang2021efficient}. For non-stabilizer states and local measurements, both classical shadows and the derandomized version exhibit a worst-case sample complexity of $\mathcal{O}(4^n/\epsilon^2)$. In contrast, standard DFE and our protocol require $\mathcal{O}(2^n/\epsilon^2)$ copies. While this scaling is an average-case analysis, by Markov’s inequality, the actual cost is unlikely to deviate significantly from the average \cite{flammia2011direct}. In the classical shadows framework with entangled (Clifford) measurements, the sample complexity becomes independent of the Hilbert space dimension. However, implementing entangled operations remains challenging and noise-prone on current quantum devices \cite{preskill2018quantumcomputingin}. Moreover, as shown in Ref.~\cite{leone2023nonstabilizerness}, the associated classical post-processing overhead can exceed the total resource cost of DFE for non-stabilizer target states. Therefore, our protocol offers favorable resource scaling among the methods discussed in this paragraph.

\section{Numerical simulations} \label{sec:numerical}

\begin{figure*}[t]
    \centering
    \begin{subfigure}{0.32\textwidth}
        \centering
        \includegraphics[width=\textwidth]{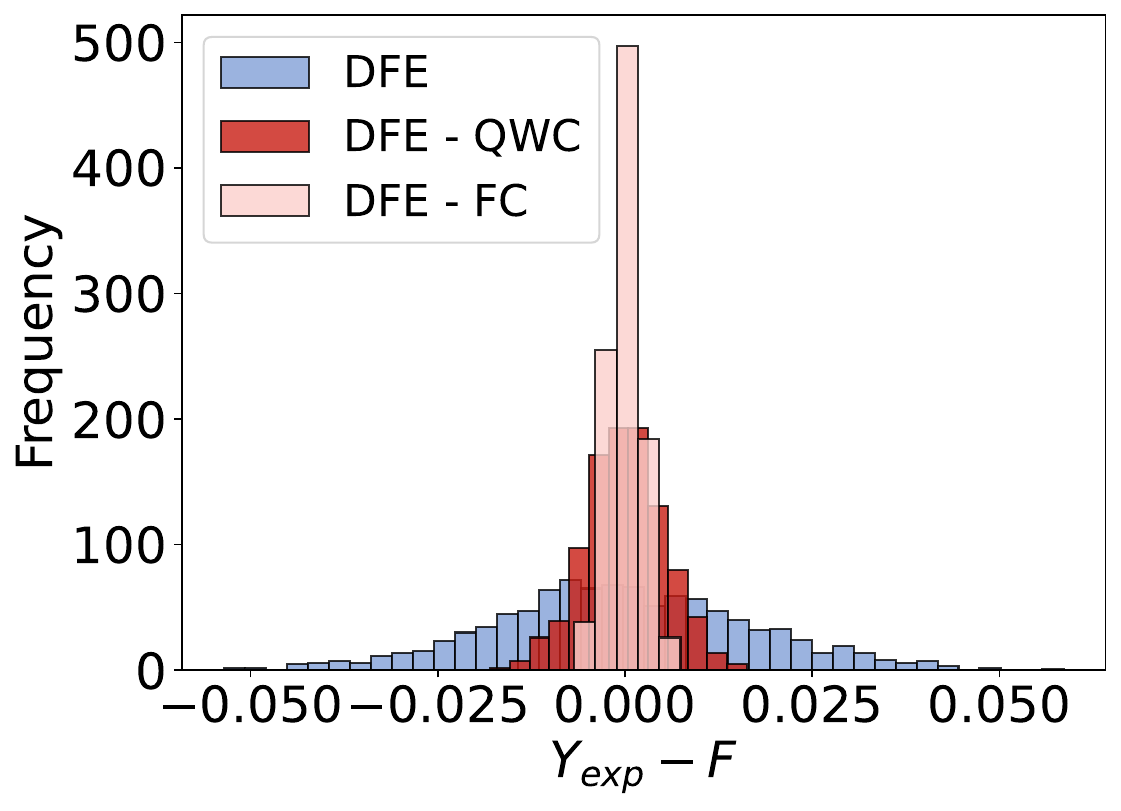}
        \caption{Haar state fidelity}\label{fig:haar_fidelity}
    \end{subfigure}
    \begin{subfigure}{0.32\textwidth}
        \centering
        \includegraphics[width=\textwidth]{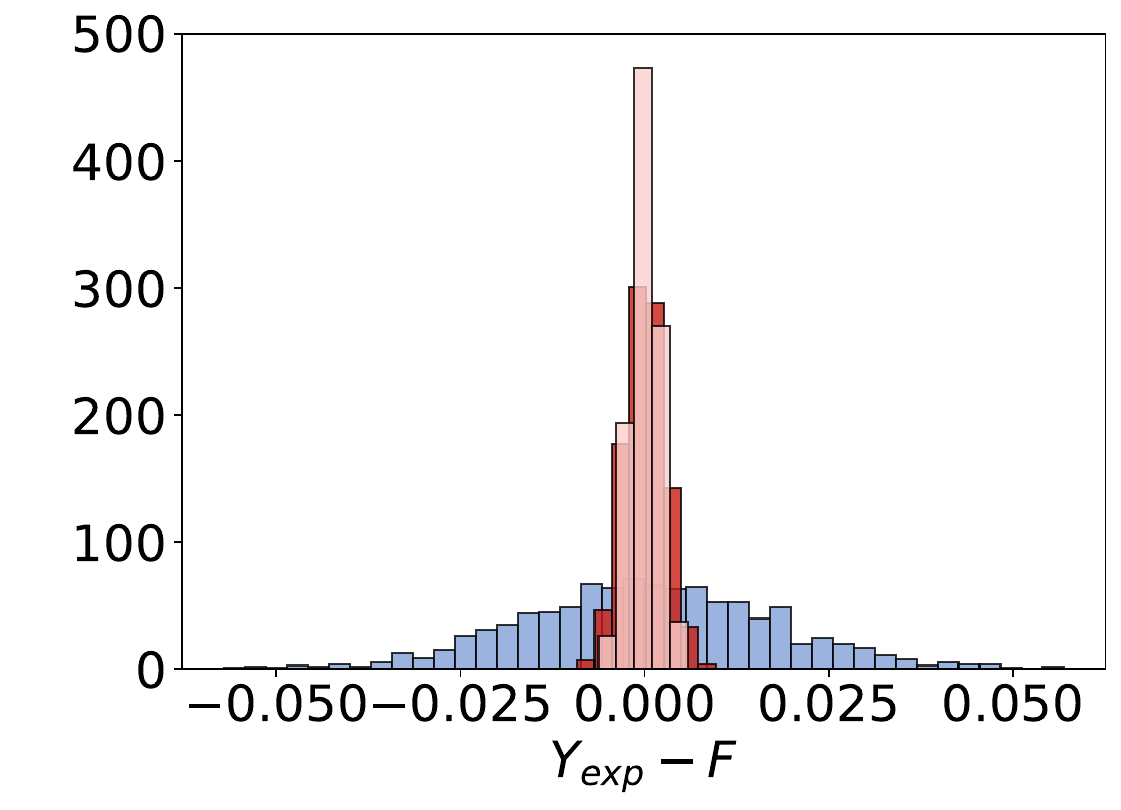}
        \caption{W state fidelity}\label{fig:w_fidelity}
    \end{subfigure}
    \begin{subfigure}{0.32\textwidth}
        \centering
        \includegraphics[width=\textwidth]{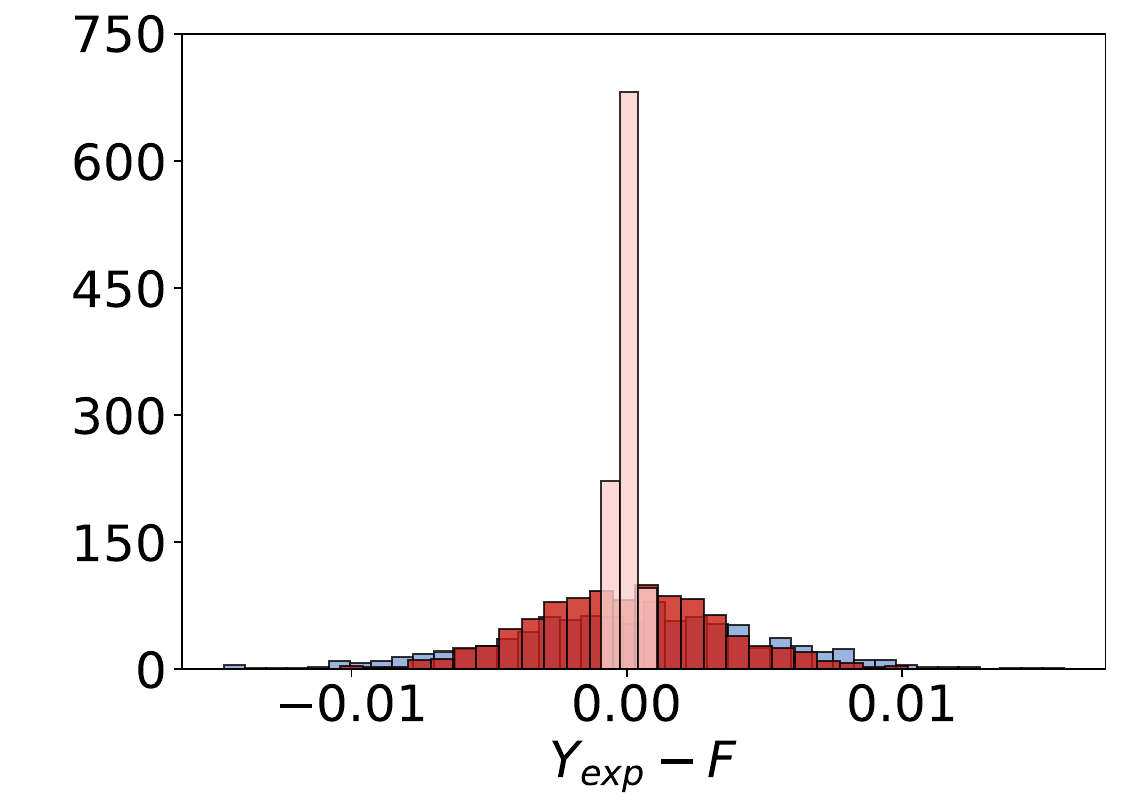}
        \caption{GHZ state fidelity}\label{fig:ghz_fidelity}
    \end{subfigure}

    \begin{subfigure}{0.32\textwidth}
        \centering
        \includegraphics[width=\textwidth]{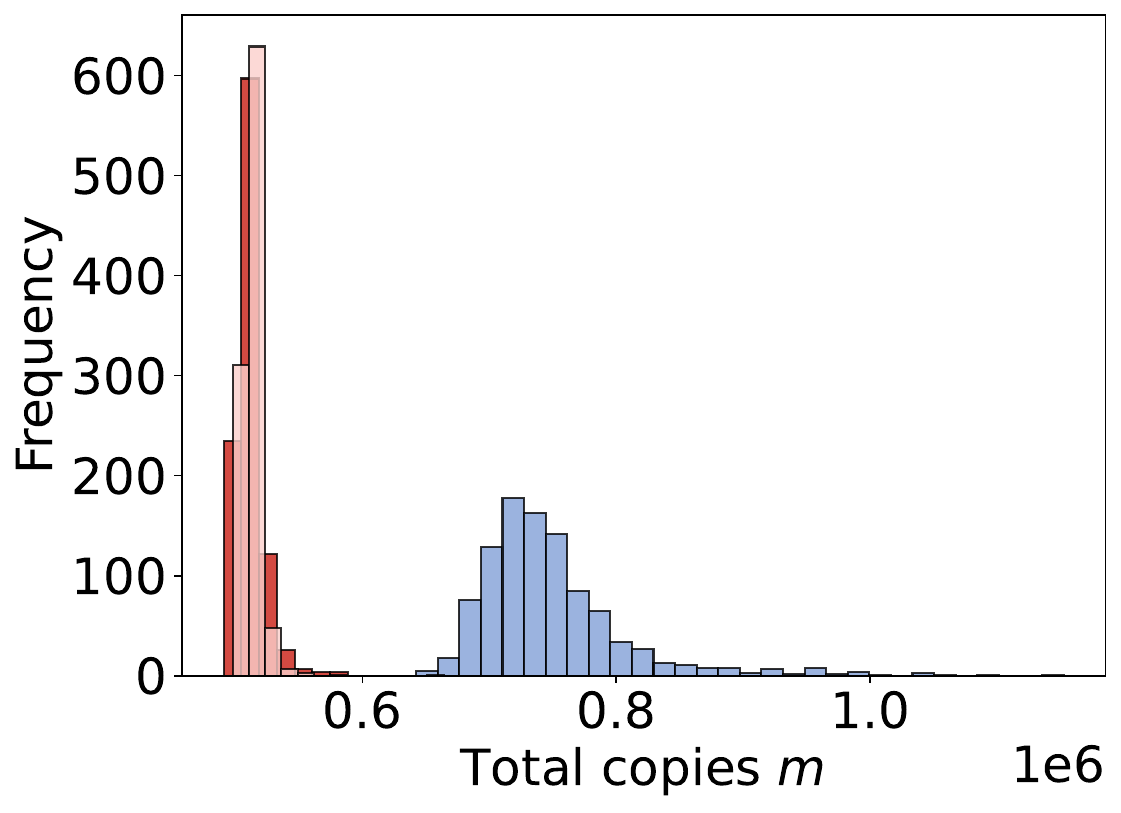}
        \caption{Haar state copies}\label{fig:haar_copies}
    \end{subfigure}
    \begin{subfigure}{0.32\textwidth}
        \centering
        \includegraphics[width=\textwidth]{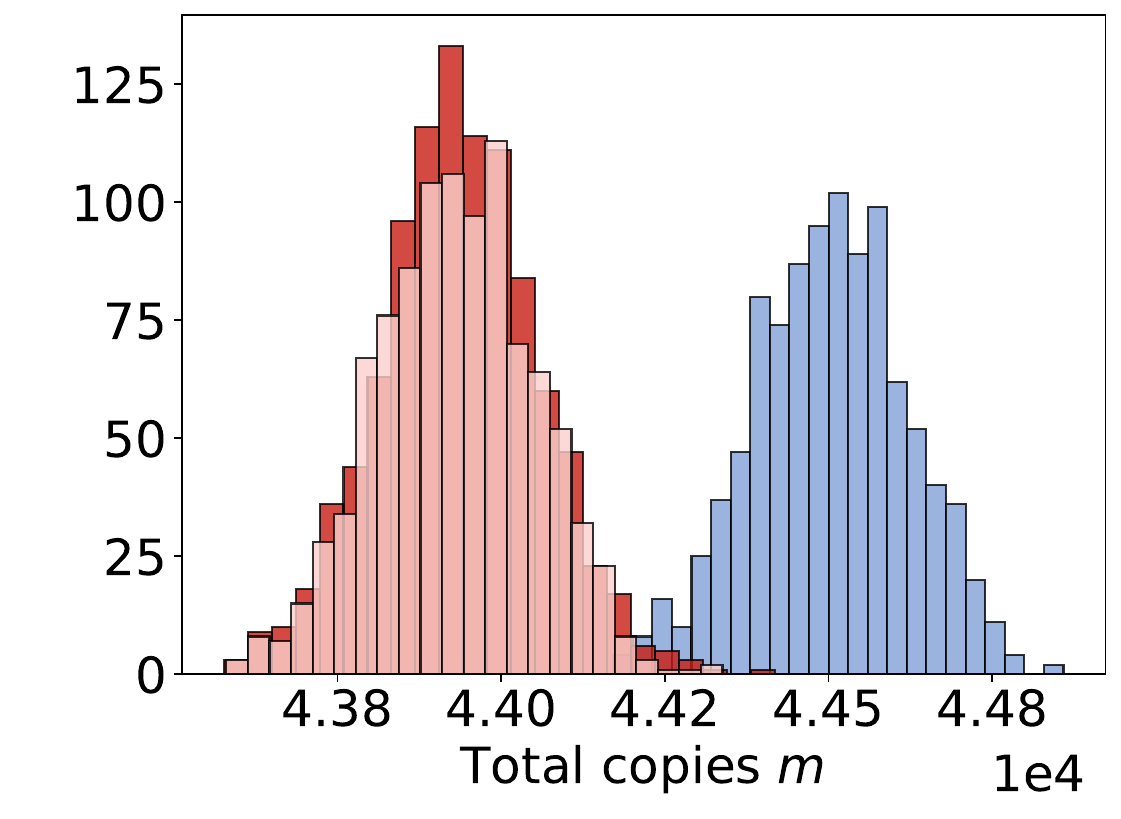}
        \caption{W state copies}\label{fig:w_copies}
    \end{subfigure}
    \begin{subfigure}{0.33\textwidth}
        \centering
        \includegraphics[width=\textwidth]{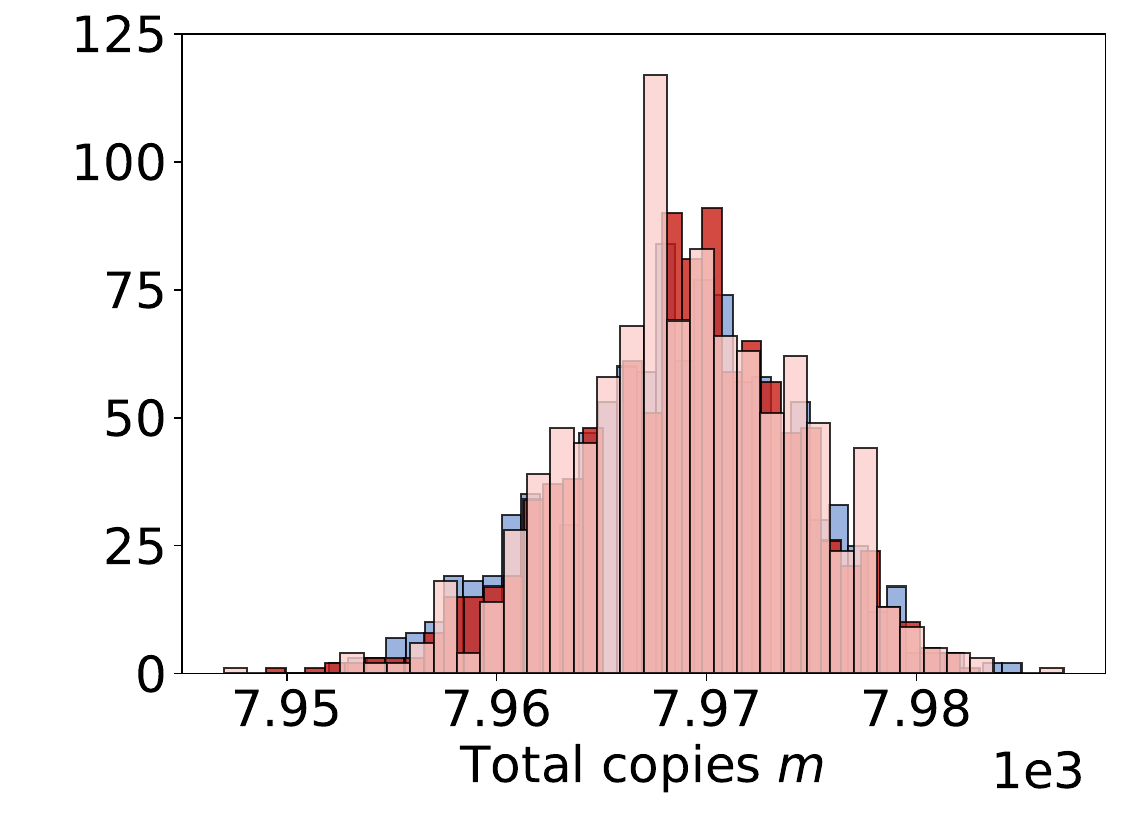}
        \caption{GHZ state copies}\label{fig:ghz_copies}
    \end{subfigure}
    \caption{\justifying The plots are obtained using 1000 samples of 8-qubit states with parameters set to $p=0.1, \varepsilon = 0.05$, $\delta = 0.05$, $\ell = 8000$. In (a), (b), and (c), we show the distribution of the residual error of the fidelity ($Y_{exp}$ - $F$), represented by the frequency of occurrence of different error values across samples. Results are shown for the DFE protocol (blue) and our method with QWC (red) and FC (blue) using Haar-random states, W states and GHZ states. Figures (d), (e), and (f) show the number of copies required by each protocol. For Haar-random states, the variance of $Y_{exp}$ is reduced by $92\%$ ($99\%$) with QWC (FC) and the mean number of copies is reduced by $32\%$. In Fig.~\ref{fig:haar_copies}, we omit the last 5 samples from the dataset used in the original DFE and QWC analysis for better visualization. }
    \label{fig:plot_8qubits}
\end{figure*}

To study the performance of the proposed method, we carried out numerical simulations and benchmarked it against DFE protocol \cite{flammia2011direct}. We tested fidelity estimation on three types of states: Haar-random states, W states, and the GHZ state. These ideal input states $\rho$ were subjected to a depolarizing channel, producing the noisy state
\begin{equation} 
\sigma := (1-p) \rho + p \frac{\mathds{1}}{d}, 
\end{equation}
where $p$ is the noise level.

Our protocol differentiates from DFE by incorporating a preprocessing step, polynomial in $d$, that involves grouping Pauli strings. To achieve this, we employed the SI algorithm, within the non-overlapping framework, for both full and qubit-wise commutation. This algorithm minimizes the estimator variance among other grouping techniques while remaining computationally efficient~\cite{yen2023deterministic}. The algorithm first sorts Pauli strings in decreasing order based on their weights, $ |b_{kl}|$, and then forms the groups (see App.~\ref{app:grouping-techniques} for the pseudocode).  This approach ensures that the largest groups consist of Pauli strings with the highest weights, thereby reducing the number of copies and the variance of the expected value of the operator under consideration.    

In Fig.~\ref{fig:plot_8qubits}, we show the distribution of the residual error of the fidelity, represented as the frequency with which different error values occur across 1000 simulated 8-qubit states. We also report the total number of copies required. Both metrics are benchmarked against the direct fidelity estimation protocol outlined in Sec.~\ref{sec:dfe}. The simulations were executed with parameters set to $p=0.1, \varepsilon = 0.05$, $\delta = 0.05$, which resulted in $\ell = 1/\varepsilon^2 \delta = 8000$. For Haar-random states, the results show that using grouping techniques for DFE significantly reduces the variance of the fidelity estimator by $92\%$ for QWC and $99\%$ for FC. Similarly, Fig.~\ref{fig:haar_copies} highlights that the number of copies required decreases with grouping techniques. Both QWC and FC frameworks achieve a $32\%$ reduction in the mean number of copies, demonstrating that grouping effectively reduces the number of copies keeping the same precision of fidelity estimation.

The results in Fig.~\ref{fig:w_fidelity} and Fig.~\ref{fig:w_copies} indicate that for the W state, the variance of $Y_{exp}$ is reduced by $97\%$ with QWC and $99\%$ with FC, while the mean number of copies decreases by $1.27\%$ compared to DFE with both grouping techniques. Finally, for the GHZ state, the variance of the fidelity estimator decreased by $48\%$ with QWC and $99\%$ with FC, as illustrated in Fig.~\ref{fig:ghz_fidelity}. However, in Fig.~\ref{fig:ghz_copies}, we see that the mean number of copies remained unchanged. These results on the number of copies align with observations in Ref. \cite{flammia2011direct}, which noted that their method was optimal when these states were used as input. 
\begin{table*}[t]
\setlength{\tabcolsep}{15pt} 
\renewcommand{\arraystretch}{1.1} 
\begin{center}
\begin{tabular}{ccccccccc}
\toprule
 \multicolumn{1}{c}{{$n$}} & \multicolumn{2}{c}{$g$} & \multicolumn{3}{c}{Variance $(1\times 10^{-5})$} & \multicolumn{3}{c}{$\mathbb{E}[m]$} \\
\cmidrule(lr){2-3}\cmidrule(lr){4-6} \cmidrule(lr){7-9}
\multicolumn{1}{c}{}& QWC & FC &DFE  & QWC & FC &DFE  & QWC & FC \\
\midrule
\multicolumn{1}{c}{2}& 10 & 7  & $12.2$  & $7.2$ &$ 4.8$ & 13803 & 12630 & 12914 \\ 

\multicolumn{1}{c}{3}& 28 & 15 & $18.4$  &  $6.2$ & $4.0$ & 26660  &  22705 &  23545 \\ 

\multicolumn{1}{c}{4}& 82 & 33  & $25.3$ & $6.3$ & $2.5$ & 50572 & 40126 & 42523 \\ 

\multicolumn{1}{c}{5}& 244 & 74  &  $26.8$ & $5.3 $& $1.8$ &   98439 &  73166 & 77357 \\ 

\multicolumn{1}{c}{6}& 730 & 172 &  $28.7$ & $4.1$ & $1.2$ & 191706  & 138117  & 142610 \\ 

\multicolumn{1}{c}{7}& 2188 & 399 &  $33.1$  & $3.3$ & $0.7$ &  380112 &  262896 &   268826 \\ 

\multicolumn{1}{c}{8}& 6562 & 924 & $33.6$ & $2.6$ & $0.4$ & 754950 &  513139 & 513995 \\ 
\bottomrule
\end{tabular}
\end{center}
\caption{\justifying Number of groups $g$ of Pauli operators in the sorted insertion (SI) method for full and qubit-wise commutativity (FC and QWC) across number of qubits $n$, along with the variance of the fidelity estimator $Y_{exp}$ and the mean number of copies $\mathbb{E}[m]$ for the original protocol and the two grouping techniques, with $m = \sum_{k_i} m_{k_i}$ and $m_{k_i}$ given in Eq.~\eqref{eq:copies-original} and Eq.~\eqref{eq:copies-our}. }
\label{table:dfe-performance}
\end{table*}

To further benchmark our protocol, we conducted experiments with 1000 samples of Haar-random states for different system dimensions, using the same simulation parameters as in Fig.~\ref{fig:plot_8qubits}: $p=0.1$, $\varepsilon = 0.05$, $\delta = 0.05$, and $\ell = 8000$. 
The data obtained is shown in Tab.~\ref{table:dfe-performance}, where we observe that the number of groups scales polynomially with $d$, with fewer groups required for the FC technique. This observation lines up with the analysis made in Ref. \cite{yen2023deterministic} that numerically demonstrates that the total number of groups $g$ under FC is significantly smaller than for QWC. We also note that the variance of $Y_{exp}$ is consistently lower when FC is employed, and this reduction increases with $d$. In particular, the variance decreases by $53\%$ and $63\%$ for smaller systems, reaching reductions of $92\%$ and $99\%$ for larger systems, for QWC and FC, respectively. Similarly, using both commutating frameworks, the reduction of the mean number of copies also grows with $d$, starting at approximately $6\%$ for $n = 2$ and reaching $33\%$ for $n = 8$. Notably, for smaller systems, QWC yields slightly greater reductions in the number of copies, which may be attributed to a more uneven distribution of the weights associated with the Pauli operators, favoring fewer copies.  

Another potential approach to reduce the number of copies is through overlapping grouping techniques~\cite{yen2023deterministic}, such as the ICS algorithm introduced in Sec. \ref{sec:grouping}. ICS forms overlapping groups based on the covariances of commuting Pauli operators but relies on the prior identification of non-overlapping groups using the SI algorithm \cite{yen2023deterministic}. However, this added classical complexity makes ICS a less efficient option for reducing resources in DFE.


\section{Conclusions \label{sec:disc}}


Given the importance of the DFE protocol, it is crucial to develop improvements that reduce its measurement cost while maintaining precision. One promising approach is to optimize the measurement strategy by grouping commuting Pauli operators, which reduces the number of distinct measurement settings and improves estimation efficiency~\cite{izmaylov2019revising, verteletskyi2020measurement, jena2022optimization, yen2020measuring, izmaylov2019unitary, zhao2020measurement, hamamura2020efficient, Crawford_2021, yen2023deterministic, hadfield2022measurements, gresch2023guaranteed, wu2023overlapped}. While such techniques have been widely explored for general observable estimation, their application to DFE has remained largely unexplored.

In this work, we integrated Pauli operator grouping techniques into the DFE protocol to reduce the total number of measurements required while maintaining precision. In our method, we group the Pauli strings appearing in the decomposition of the target state into commuting families and perform weighted sampling over these groups. The expectation values of these operators are then estimated by measuring in the common eigenbases of each group. We proved that, for a fixed error $\epsilon$ and confidence $1-\delta$, the total number of copies required by our protocol is upper-bounded by that of the standard DFE protocol. Furthermore, we also upper-bounded the variance of our protocol by that of standard DFE under certain conditions.


Numerical simulations on 8-qubit Haar-random states confirmed the theoretical predictions. For local measurements, our approach achieved a $92\%$ reduction in variance and a $32\%$ reduction in the number of copies required compared to the standard DFE protocol. Additional reductions in variance were achieved when allowing for entangled measurements, though these come with higher implementation costs that are often avoided in practice.


While our protocol and simulations assume ideal measurements, experimental imperfections, such as gate and readout errors, can introduce bias and lead to systematic deviations in fidelity estimates. Fortunately, the protocol, when implemented within the QWC framework, relies only on single-qubit gates, which are among the most accurate operations available on current quantum hardware \cite{barends2014superconducting, smith2025single}. Although readout errors remain a relevant source of noise, they could be mitigated by incorporating existing error mitigation techniques into our protocol \cite{chen2019detector, maciejewski2020mitigationofreadout, bravyi2021mitigating}. Moreover, recent advances in error mitigation for randomized measurement protocols~\cite{seshadri2024theory,chen2021robust, hu2025demonstration} may be adaptable to our setting, potentially further enhancing robustness. Investigating these directions could yield improved practical performance and pave the way for more noise-resilient implementations of our protocol.


\section*{Code availability}
Simulations were run using \texttt{Pyhton} 3.11.9 on an Intel Core i5-7500 3.4 GHz 6 MB L3 cache. The computations could be optimized by utilizing a GPU or parallelization.
The scripts to run the experiments presented in this work can be found in GitHub in the repository \texttt{\href{https://github.com/juliabarbera/DFE-sampling-pauli-groups}{juliabarbera/DFE-sampling-pauli-groups}}.

\begin{acknowledgments}
This work was supported by the Government of Spain (Severo Ochoa CEX2019-000910-S, FUNQIP, Misiones CUCO Grant MIG-20211005, European Union NextGenerationEU PRTR-C17.I1 and Quantum Spain), Fundació Cellex, Fundació Mir-Puig and Generalitat de Catalunya (CERCA program and Grup reconegut 2021 SGR 01440). J.B.R. has received funding from the “Secretaria d’Universitats i Recerca del Departament de Recerca i Universitats de la Generalitat de Catalunya” under grant FI-1 00096, as well as the European Social Fund Plus. M.N. acknowledges funding from the European Union’s Horizon Europe research and innovation programme under the Marie Skłodowska-Curie Grant Agreement No. 101081441.
\end{acknowledgments}

\bibliographystyle{apsrev4-2}
\bibliography{bib}
\newpage
\appendix
\onecolumngrid
\section{Sorted insertion}\label{app:grouping-techniques}
In this section, we present the pseudocode for the sorted insertion (SI) algorithm introduced in Ref.~\cite{Crawford_2021}. The SI algorithm is designed to group Pauli strings into commuting families while considering their variances. The goal is to minimize the total number of measurements required and ensure low variances in the resulting estimators.

\hspace{2mm}

\begin{algorithm}[h]
\caption{SI Algorithm}
\KwIn{
    \texttt{pauli\_list}: List of strings representing the Pauli operators,
    \texttt{x\_ki\_list}: List of coefficients corresponding to the Pauli operators,
     \texttt{condition}: Specifies the commutation framework, which can be either fully commuting or qubit-wise commuting.
    
}
\KwOut{\texttt{no\_groups}: List of groups, each containing commuting Pauli operators, \texttt{x\_ki\_no\_groups}: A list of weights grouped in correspondence with the Pauli strings}

\BlankLine

\textbf{Initialize:} \texttt{no\_groups} $\gets \emptyset$, \texttt{x\_ki\_no\_groups} $\gets \emptyset$\\
Sort \texttt{pauli\_list} in decreasing order based on the absolute values of their corresponding weights, \texttt{x\_ki\_list}\\
\ForEach{\texttt{pauli} in \texttt{pauli\_list} with index $i$}{
    \texttt{added\_to\_any\_group} $\gets$ \texttt{False}
    
    \ForEach{\texttt{group, x\_ki\_group} in \texttt{zip(no\_groups, x\_ki\_no\_groups)}}{
    
        \If{\texttt{commutes(pauli, group, condition)}}{
            Append \texttt{pauli} to \texttt{group} \\
            Append \texttt{x\_ki\_list[i]} to \texttt{x\_ki\_group} \\
            \texttt{added\_to\_any\_group} $\gets$ \texttt{True} \\
            \textbf{break}
        }
    }
    \If{\textbf{not} \texttt{added\_to\_any\_group}}{
        Append \texttt{[pauli]} to \texttt{no\_groups} \\
        Append \texttt{[x\_ki\_list[i]]} to \texttt{x\_ki\_no\_group}
    }
}

\Return{\texttt{no\_groups}, \texttt{x\_ki\_no\_group}}
\end{algorithm}


\onecolumngrid
\section{Variance of the original and grouped protocols}\label{app:variance-bound}
In this section, we demonstrate that, under the approximation $\lceil m_{k} \rceil \approx m_{k}$, the variance of our protocol is bounded by the variance of the original protocol. This suggests that our approach provides an improvement over the original method in terms of variance. The approximation holds when $\Vert \mathbf{b}_{k}\Vert_1^2 / \Vert \mathbf{b}_{k}\Vert^4$ is sufficiently large, which is the case for Haar-random states. In such states, the coefficients $|b_k|$ are approximately $1/d$ for large $d$. 

Let $\hat{Y}$ and $\tilde{Y}$ represent the fidelity estimators of DFE and our protocol, respectively. The objective is to show that $\mathrm{Var}(\tilde{Y}) \leq \mathrm{Var}(\hat{Y})$. Since both estimators are sums of independent and identically distributed random variables $\hat{X}_{k_i}$ and $\tilde{X}_{k_i}$, this is equivalent to proving that $\mathrm{Var}(\tilde{X}_{k_i}) \leq \mathrm{Var}(\hat{X}_{k_i})$. Furthermore, as the expected value of both estimators equals the theoretical fidelity (i.e., $\mathbb{E}[\tilde{X}_{k_i}]^2 = \mathbb{E}[\hat{X}_{k_i}]^2 = \tr{\rho\sigma}^2$), the problem reduces to showing that $\mathbb{E}[\tilde{X}_{k_i}^2] \leq \mathbb{E}[\hat{X}_{k_i}^2]$. Let $k_i$ be the index of the Pauli string measured in experiment round $i$, and $r = r_{1} \dots r_{m_{k_i}}$ specifies which one of the $d$ eigenstates of $\hat{X}_{k_i}$, given the Pauli string $k_i$, is obtained for each of the $m_{k_i}$  measurement shots. 
Since for each $\hat{X}_{k_i}$ we are considering $m_{k_i}$ shots, the random variables $\hat{X}_{k_i}$ in the DFE protocol follow the probability distribution
\begin{align}
    p(k_i, r) = p(k_i) p(r|k_i) = p(k_i) p(r_1|k_i) \dots p(r_{m_{k_i}}|k_i)  = b_{k_i}^2 \langle r_1| \rho |r_{1}\rangle \dots \langle r_{m_{k_i}}| \rho |r_{m_{k_i}}\rangle.
\end{align}
We start analyzing the squared expected value of the estimator in the original protocol
\begin{align}
    \mathop{\mathbb{E}}_{k_i,  r}[\hat{X}_{k_i}^2] &= \mathop{\mathbb{E}}_{ k_i} \mathop{\mathbb{E}}_{r|k_i}  \left[\left(\frac{1}{m_{k_i} b_{k_i} \sqrt{d}} \sum_{j=1}^{m_{k_i}} c_{k_i r_j}\right)^2\right]e= \mathop{\mathbb{E}}_{ k_i} \mathop{\mathbb{E}}_{r|k_i} \left[\frac{1}{m^2_{k_i} b^2_{k_i} d}\left( \sum_{j=1}^{m_{k_i}} c_{k_i r_j}^2 + \sum_{j \neq l}^{m_{k_i}} c_{k_i r_j} c_{k_i r_l} \right)\right].
\end{align}
The first term in the last line simplifies to $m_{k_i}$, since $c_{k_i r_j} \in \{-1,1\}$. Calculating the conditional expectation, the second term can be rewritten as $\sum_{j \neq l} \mathop{\mathbb{E}}_{ r|k_i} [c_{k_i r_j} c_{k_i r_l}] = \sum_{j \neq l} \mathop{\mathbb{E}}_{ r_j|k_i} [c_{k_i r_j}] \mathop{\mathbb{E}}_{ r_l|k_i} [c_{k_i r_l}]$, since the $j$-th and $l$-th samples are independent conditioned on $k_i$. Also, 
\begin{align}
    \mathop{\mathbb{E}}_{r_j |k_i} [c_{k_i r_j}] = \sum_{r_j=1}^d c_{k_i r_j}  \langle r_{j}| \rho |r_{j}\rangle = a_{k_i} \sqrt{d}. 
\end{align}
Thus, we have
\begin{align}\label{eq:upper-bound-f-dfe}
     \mathop{\mathbb{E}}_{ k_i, r}[\hat{X}_{k_i}^2] &=  \mathop{\mathbb{E}}_{ k_i}\left[\frac{1}{m_{k_i} b^2_{k_i} d}\right] +    \mathop{\mathbb{E}}_{ k_i}\left[\frac{1}{m_{k_i}^2 b^2_{k_i} d}\sum_{j\neq l}^{m_{k_i}} \mathop{\mathbb{E}}_{ r_{j}|k_i}\left[c_{k_i r_j}\right] \mathop{\mathbb{E}}_{ r_{l}|k_i}\left[c_{k_i r_l}\right]\right] \nonumber\\
     &= \mathop{\mathbb{E}}_{k_i}\left[\frac{1}{m_{k_i} b^2_{k_i} d}\right] +   \mathop{\mathbb{E}}_{k_i}\left[\frac{1}{m_{k_i}^2 b^2_{k_i}}\sum_{j\neq l}^{m_{k_i}} a_{k_i}^2 \right] \nonumber\\
    &= \mathop{\mathbb{E}}_{k_i}\left[\frac{1}{m_{k_i} b^2_{k_i} d}\right] +   \mathop{\mathbb{E}}_{k_i}\left[\frac{1}{m_{k_i} b^2_{k_i}} (m_{k_i}-1)a_{k_i}^2 \right] \nonumber\\
    &\approx \frac{\ell \epsilon^2}{2 \ln (2/\delta)} \left(\sum_{k_i}b_{k_i}^2 - d\sum_{k_i}b_{k_i}^2a_{k_i}^2\right) + \sum_{k_i}a_{k_i}^2. 
\end{align}
Here, we use the fact that the number of copies in the original protocol is given by Eq.~\eqref{eq:copies-original}, and assume the approximation $\lceil m_{k_i} \rceil \approx m_{k_i}$ holds. 

We now compute the same quantity for our protocol. The calculation follows a similar approach, but with the key difference that the sampling is performed over groups of Pauli operators. Here, $k_i$ corresponds to the selected group, and $c_{k_i r_j} = \sum_{l} c_{k_i l}^{(r_j)} b_{k_il}$, where the sum is taken over the elements in the group, with $b_{k_i l}$ representing the coefficient associated to the Pauli $l$ of the group $g_k$. One result we need is $\mathop{\mathbb{E}}_{r_j | k_i } [c_{k_il}^{(r_j)}] = a_{k_i l}$. Then, we have that
\begin{align}\label{eq:upper-bound-f-ours}
    \mathop{\mathbb{E}}_{k_i,  r}[\tilde{X}_{k_i}^2] &= \mathop{\mathbb{E}}_{k_i,  r}\left[\left(\frac{1}{m_{k_i}\sqrt{d}\lVert \mathbf{b}_{k_i} \rVert^2 } \sum_{j=1}^{m_{k_i}} \sum_{l = 1}^{g_k} c_{k_i l}^{(r_j)} b_{k_i l}\right)^2\right] \nonumber \\
    &= \mathop{\mathbb{E}}_{k_i}\left[\frac{1}{m_{k_i}^2 d \lVert \mathbf{b}_{k_i} \rVert^4 } \left(\sum_{j=1}^{m_{k_i}} \sum_{l =1}^{g_k} \mathop{\mathbb{E}}_{ r_{j}|k_i} \left[\left(c_{k_i l}^{(r_j)} b_{k_il}\right)^2\right] + \sum_{j\neq l = 1}^{m_{k_i}} \sum_{l, t = 1}^{g_k} \mathop{\mathbb{E}}_{ r_{j}|k_i} \left[c_{k_i l}^{(r_j)} \right] \mathop{\mathbb{E}}_{ r_{l}|k_i} \left[c_{k_i t}^{(r_l)}\right] b_{k_il} b_{k_it} \right)\right] \nonumber \\
    & = \mathop{\mathbb{E}}_{k_i}\left[\frac{1}{m_{k_i} d \lVert \mathbf{b}_{k_i} \rVert^4} \sum_{l = 1}^{ g_k} b_{k_il}^2\right] +   \mathop{\mathbb{E}}_{k_i}\left[\frac{1}{m_{k_i}^2 \lVert \mathbf{b}_{k_i} \rVert^4}\sum_{j \neq l}^{m_{k_i}} \sum_{l,t =1}^{g_k} a_{k_il} a_{k_it} b_{k_il} b_{k_it} \right]\nonumber \\
    & \leq  \mathop{\mathbb{E}}_{k_i}\left[\frac{1}{m_{k_i} d \lVert \mathbf{b}_{k_i} \rVert^2}\right] +   \mathop{\mathbb{E}}_{k_i}\left[\frac{1}{m_{k_i} \lVert \mathbf{b}_{k_i} \rVert^2} (m_{k_i}-1) ||\mathbf{a}_{k_i}||^2 \right] \nonumber\\
    & \leq \frac{\ell \epsilon^2}{2 \ln (2/\delta)} \left(\sum_{k_i}||\mathbf{b}_{k_i}||^2 - d\sum_{k_i}||\mathbf{b}_{k_i}||^2||\mathbf{a}_{k_i}||^2 \right) + \sum_{k_i}||\mathbf{a}_{k_i}||^2 \nonumber\\
    & \leq \frac{\ell \epsilon^2}{2 \ln (2/\delta)} \left(\sum_{k_i, l}{b}_{k_i l}^2 - d\sum_{k_i, l} {b}_{k_i l}^2 {a}_{k_i l}^2 \right) + \sum_{k_i, l}{a}_{k_i l}^2.
\end{align}
In the fourth line, we used the definition of $||\mathbf{b}_{k_i}||^2$ from Eq.~\eqref{eq: new prob dist} for the first term, and applied the Cauchy-Schwarz inequality, $\left(\sum_{l =1}^{g_k} a_{k_il} b_{k_il}\right)^2 \leq ||\mathbf{b}_{k_i}||^2|| \mathbf{a}_{k_i}||^2$, for the second term. In the fifth line, we substituted the expression for the number of copies $m_{k_i}$ derived in Eq.~\eqref{eq:copies-our}, considering the approximation $\lceil m_{k_i} \rceil \approx m_{k_i}$. Additionally, we utilized the inequality $||\mathbf{b}_{k_i}||_2 \leq ||\mathbf{b}_{k_i}||_1$. 
From Eq.\eqref{eq:upper-bound-f-ours}, it follows that the variance of our protocol is upper-bounded by the variance of the standard DFE protocol in the limit $m_{k_i} \gg 1$.

\end{document}